\documentclass{LT23auth}
\usepackage{graphicx}

\begin{document}

\begin{frontmatter}

\title{Infrared properties of W-doped charge-density-wave material
K$_{0.3}$MoO$_3$}

\author[address1]{T. Feng},
\author[address1]{N. L. Wang\thanksref{thank1}},
\author[address1]{P. Zheng},
\author[address1]{Z. J. Chen},
\author[address2]{M. L. Tian}

\address[address1]{Institute of Physics, Chinese Academy of Sciences, P. O. Box 603, Beijing
100080, PR China}

\address[address2]{Structure Research Laboratory, University of Science and Technology of China, Hefei 230026, P. R. China}

\thanks[thank1]{ E-mail:nlwang@cl.cryo.ac.cn}

\begin{abstract}
The optical conductivity spectra of quasi-one dimensional
compounds K$_{0.3}$Mo$_{1-x}$W$_x$O$_3$ (x=0, 0.03 and 0.15) have
been studied over broad frequencies. While the dc resistivity
measurements indicate no sign of CDW transition in heavily W-doped
blue bronze, the optical conductivity spectra still show a single
particle gap at around 0.2 eV for \textbf{E} parallel to the chain
direction. Such impurity effect challenges our understanding about
the occurrence of the optical gap with the CDW transition.
\end{abstract}

%
%
\end{frontmatter}
A central aspect of the charge-density-wave (CDW) phase transition
is the appearance of a single-particle gap, which is usually
considered as a typical feature associated with CDW condensate.
However, it is found that the gap, for example in
K$_{0.3}$MoO$_3$, could exist at temperatures much higher than the
transition temperature.\cite{Degiorgi,Schwartz} Although
mean-field theory predicted that the absorption is zero for
frequency less than the gap, 2$\Delta$, and contains an
inverse-square-root singularity at
$\omega$=2$\Delta$,\cite{Schulz} the experimental observation is
quite different. The singularity is absent, and there is a
substantial tail below the maximum. Some theories explain those as
due to the zero-point and the thermal lattice fluctuation
effects,\cite{Degiorgi,Schwartz,McKenzie} but a full understanding
of these phenomena still requires more efforts in both experiments
and theories. Impurities doping is an important technique for
studying the mechanism of CDW. It would be very interesting to see
how impurities affect the single-particle gap seen in infrared
experiment. In this work, we study the pure and tungsten-doped
K$_{0.3}$MoO$_3$ CDW materials. The experiment seems to indicate
that the observed optical gap is irrelevant to the Peierls
transition.

Pure and W-doped K$_{0.3}$MoO$_3$ crystals were grown by
electrolytic reduction of a molten mixture of K$_2$MoO$_4$ and
MoO$_3$. The dc conductivity was measured by a standard four-probe
technique. The reflectivity data at various temperatures were
collected in a Bruker 66v/s spectrometer with light polarization
parallel to the chain direction in a broad frequency range from 50
cm$^{-1}$ to 30000 cm$^{-1}$. The optical conductivity spectra
were extracted from the Kramers-Kroning transformation of the
reflectivity.

Figure 1 shows dc conductivity of K$_{0.3}$MoO$_3$,
K$_{0.3}$Mo$_{0.97}$W$_{0.03}$O$_3$ and
K$_{0.3}$Mo$_{0.85}$W$_{0.15}$O$_3$. The pure K$_{0.3}$MoO$_3$
undergoes a clear metal-insulator (M-I) transition at around 180K.
The conductivity increases with decreasing temperature at high
temperature, but decreases sharply below 180 K. For
$_{0.3}$Mo$_{0.97}$W$_{0.03}$O$_3$, the M-I transition occurs in a
broad temperature region and the conductivity changes gradually.
Therefore, 3\% W doping smears the Peierls transition.
Conductivity of K$_{0.3}$Mo$_{0.85}$W$_{0.15}$O$_3$ decreases
monotonously from room temperature to low temperature, which is a
typical insulating-like behavior. In this case there is no
evidence of Peierls transition and CDW is completely suppressed by
15\% W impurities.

\begin{figure}[t]
\begin{center}\leavevmode
\includegraphics[width=1.0\linewidth]{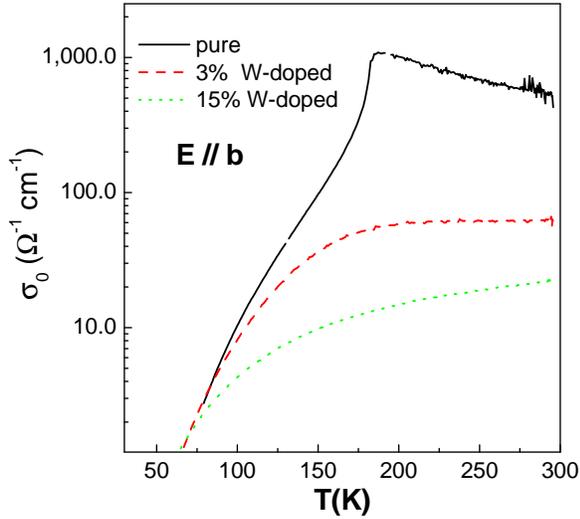}
\caption{The temperature-dependent dc conductivity for
K$_{0.3}$MoO$_3$, K$_{0.3}$Mo$_{0.97}$W$_{0.03}$O$_3$ and
K$_{0.3}$Mo$_{0.85}$W$_{0.15}$O$_3$ along b-direction.
}\label{figurename}\end{center}\end{figure}

Figure 2 shows the optical conductivity spectra of
K$_{0.3}$MoO$_3$, K$_{0.3}$Mo$_{0.97}$W$_{0.03}$O$_3$ and
K$_{0.3}$Mo$_{0.85}$W$_{0.15}$O$_3$ at 77K in the polarization of
$\textbf{E}\parallel$b-axis. In mid-infrared region the raw
reflectivity data (not shown) decreases with increasing dopant,
consequently conductivity decreases with W in this region. Below
700 cm$^{-1}$, many sharp peaks appear in reflectance and
conductivity spectra of K$_{0.3}$MoO$_3$. These absorptions are
assigned to the collective modes of distorted lattice vibrations
being coupled with CDW condensate--so called
phase-phonons.\cite{Degiorgi1,Degiorgi} Similar peaks also exist
in the conductivity spectra of K$_{0.3}$Mo$_{0.97}$W$_{0.03}$O$_3$
and K$_{0.3}$Mo$_{0.85}$W$_{0.15}$O$_3$. But the number and the
amplitude of these peaks are substantially reduced. In addition,
the central frequencies of those peaks are quite different from
those in pure K$_{0.3}$MoO$_3$. This suggests that W impurities
affect the symmetry of lattice, which result in a change of phase
phonon mode.

Another striking feature is the suppression of the Low-$\omega$
spectral weight in conductivity spectrum, which gives rise to a
broad maximum at 1600 cm $^{-1}$ (0.2 eV). This absorption at 0.2
eV was assigned to the CDW single-particle gap absorption
2$\Delta$.\cite{Degiorgi,Gruner} Such gap feature, while being
most prominent in the pure sample, still exists in the two W-doped
samples and appears at the same frequency. Although the peak
position is in fair agreement with the gap value estimated from
the activation energy of dc conductivity measurement for the pure
sample (about 0.18 eV), it is quite different for the two W-doped
samples. In fact, the activated formula could not well describe
the dc conductivity curve for the W-doped samples. The
temperature-dependent dc conductivity has a tendency to evolve
towards a logarithmic behavior at high impurity concentration.
Therefore, the observed optical gap is in contradictory to the dc
result. Furthermore, as inferred from dc conductivity, CDW can not
develop in 15\% W-doped sample. Such impurity effect appears
against the scenario that the occurrence of the gap is associated
with the CDW transition.

\begin{figure}[t]
\begin{center}\leavevmode
\includegraphics[width=1.0\linewidth]{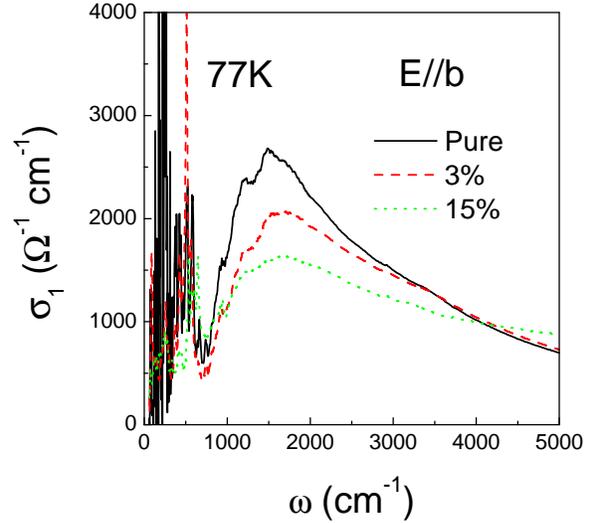}
\caption{The optical conductivity spectra of K$_{0.3}$MoO$_3$,
K$_{0.3}$Mo$_{0.97}$W$_{0.03}$O$_3$ and
K$_{0.3}$Mo$_{0.85}$W$_{0.15}$O$_3$ for
\textbf{E}$\parallel$b-axis at 77 K.
}\label{figurename}\end{center}\end{figure}

In conclusion, our dc conductivity measurements reveal the
suppression effect of W impurities on CDW transition in
K$_{0.3}$MoO$_3$. But the optical conductivity spectra still show
a single particle gap feature at around 0.2 eV for \textbf{E}
parallel to the chain direction for W-doped samples, with the
shape and position quite similar to the pure K$_{0.3}$MoO$_3$. The
observation indicates that the relation between the gap and CDW
transition is not obvious and it needs further study.

This work was supported by National Science Foundation of China
(No.10025418).

%
%

\end{document}